%% file: main.tex
\documentclass[runningheads]{llncs}
\usepackage[T1]{fontenc}
\usepackage{booktabs}
\usepackage{graphicx}
\usepackage{amsmath}
\usepackage{amsfonts}
\usepackage{amssymb}
\usepackage{xspace}
\usepackage{caption}
\usepackage{multicol}
\usepackage{multirow}
\usepackage{color}
\usepackage{hyperref}

\usepackage[show]{chato-notes}

\input{macros}

\begin{document}
%\title{\kannolo: a Sweet Library for Approximate $k$-Nearest Neighbors Search}
\title{\kannolo: Sweet and Smooth Approximate $k$-Nearest Neighbors Search}
\author{
Leonardo Delfino\inst{4} \and
Domenico Erriquez\inst{4} \and
Silvio Martinico\inst{3,1} \and\\
Franco Maria Nardini\inst{1}\orcidID{0000-0003-3183-334X} \and\\
Cosimo Rulli\inst{1}\orcidID{0000-0003-0194-361X} \and\\
Rossano Venturini\inst{2,1}\orcidID{0000-0002-9830-3936}
}
\authorrunning{L. Delfino \emph{et al.}}
\institute{ISTI-CNR, Pisa, Italy,
\email{\{name.surname\}@isti.cnr.it} \and
University of Pisa, Italy,
\email{rossano.venturini@unipi.it} \and 
University of Pisa, Italy,
\email{silvio.martinico@phd.unipi.it} 
\and University of Pisa, Italy,
\email{\{l.delfino1, d.erriquez1\}@studenti.unipi.it} 
}

\maketitle
\begin{abstract}
Approximate Nearest Neighbors (ANN) search is a crucial task in several applications like recommender systems and information retrieval. Current state-of-the-art ANN libraries, although being perfor- mance-oriented, often lack modularity and ease of use. This translates into them not being fully suitable for easy prototyping and testing of research ideas, an important feature to enable. We address these limitations by introducing \kannolo, a novel---research-oriented---ANN library written in Rust and explicitly designed to combine usability with performance effectively. \kannolo introduces a fully composable architecture for ANN search that supports both dense and sparse vector representations. It enables researchers to seamlessly mix and match different similarity measures, vector quantization techniques (e.g., Product Quantization), and index structures (e.g., HNSW) within a single unified framework. These functionalities are managed through Rust traits, allowing shared behaviors to be handled abstractly. This abstraction ensures flexibility and facilitates an easy integration of new components. In this work, we detail the architecture of \kannolo and demonstrate that its flexibility does not compromise performance. The experimental analysis shows that \kannolo achieves state-of-the-art performance in terms of speed-accuracy trade-off while allowing fast and easy prototyping, thus making \kannolo a valuable tool for advancing ANN research. Source code available on GitHub: \href{https://github.com/TusKANNy/kannolo}{\texttt{https://github.com/TusKANNy/kannolo}}.


\end{abstract}

\section{Introduction}
Nearest Neighbor (NN) search is a fundamental problem in many domains of computer science, including image processing, information retrieval, and recommendation systems. NN algorithms search through a dataset $X$ of $d$-dimensional vectors to identify the $k$ nearest neighbors to a given query point $x \in \mathbb{R}^d$. While exact solutions to this problem are computationally feasible in low-dimensional spaces, exact nearest neighbors search algorithms are no better than brute-force linear scans in high-dimensional settings.
Approximate Nearest Neighbors (ANN) techniques tackle this aspect by giving up on exactness to offer a valuable trade-off between accuracy and efficiency. These techniques are especially popular nowadays due to the advent of effective learned vector representations (embeddings) powered by Large Language Models. In response to this trend, several well-known ANN libraries have been developed.

Many of these libraries, such as NMSlib \cite{nmslib}, DiskANN \cite{diskann} (Microsoft), and ScaNN \cite{scann1} \cite{scann2} (Google) implement highly efficient ANN search algorithms optimized for speed and scalability. However, these solutions are typically built around specific indexing or quantization techniques, with rigid codebases that do not support easy integration of new components or generalization beyond their original design. This lack of flexibility poses a significant limitation for researchers and developers who need to prototype or modify ANN algorithms. Even FAISS \cite{douze2024faiss}, Meta's comprehensive library for ANN search which provides a wide variety of indexing techniques and quantization methods, lacks the modularity required for advancing research in the field. FAISS is heavily engineered for performance, leading to specialized implementations of each ANN/quantization technique. For example, in the case of the IVF (Inverted File Index), FAISS provides ten distinct classes that implement several versions of IVF index coupled with different quantizers. This highly specific architecture makes it challenging to introduce new functionalities that apply across different IVF variants, limiting flexibility in experimental settings. Moreover, FAISS suffers from a lack of support for sparse datasets and ---like other more specialized libraries--- it lacks comprehensive and easily accessible documentation. This can make it challenging to quickly understand the codebase and the functioning of the library.

In this work, we introduce \kannolo, a new Rust library for ANN. Unlike existing libraries, \kannolo is designed having ANN researchers in mind. We design it by prioritizing the ease of modification and prototyping while also supporting generality. Indeed---and differently from available competitors---\kannolo implements state-of-the-art indexing techniques for both dense and sparse vectors, as well as quantization techniques. Specifically, we design \kannolo as a collection of abstract key components such as: 1) the dataset representation (dense/sparse), 2) the quantization method, and 3) the distance/similarity measure. Second, we exploit Rust traits ---a mechanism for specifying shared behavior across types, defining a set of methods that implementing types must provide---, to implement the design above in a way that fully matches the potential unlocked by the abstraction. This modular design allows users to easily develop and integrate new indexes/quantizers and test them with minimal effort so to greatly simplify prototyping and experimentation in ANN research. Although being designed for research, \kannolo does not give up on performance. Through comprehensive benchmarks on two publicly available datasets, \sift and \msmarco, we show that \kannolo performs on par or better with the state-of-the-art competitors on both dense and sparse data, with a peak  to 11.1$\times$ and 2.1$\times$ speedups over the best competitors respectively on dense and sparse data, establishing itself as one of the most competitive ANN search libraries. 

\section{Design of \kannolo}
\kannolo is built with a modular architecture, consisting of four main components:
\begin{itemize}
\item \textbf{One-Dimensional Arrays}: The \texttt{DArray1} trait provides a unified interface for dense and sparse vectors. This abstraction allows us to build indexes and perform ANN search independently on the kind of vectors we use.
\item \textbf{Quantizer}: Quantizers transform high-dimensional data into more compact representations. They are abstracted through a \texttt{Quantizer} trait, enabling flexibility in the type of quantization used. \kannolo also supports an \textit{Identity quantizer} which leaves the data unchanged, providing a standard interface regardless of whether any vector quantization method is applied.
\item \textbf{Query Evaluator}: The \texttt{QueryEvaluator} trait is responsible for computing distances (or similarities) between dataset items and query points, and for ranking the results accordingly. It is tightly integrated with the quantizer, ensuring that all distance computations are conducted through a unified interface. This design abstracts the complexity of handling different data representations or quantization methods.
\item \textbf{Dataset}: The \texttt{Dataset} trait represents a collection of one-dimensional arrays equipped with a quantizer. It dynamically integrates with a query evaluator during search operations and provides access to the data by masking the details on the kind of vectors it contains.
\end{itemize}

To better understand how these components interact, we illustrated the indexing pipeline of \kannolo in Figure \ref{fig:indexing} and the retrieval process in Figure \ref{fig:retrieval}.
\smallbreak
\vspace{1mm}
\begin{minipage}[t]{0.46\textwidth}
\includegraphics[width=\textwidth]{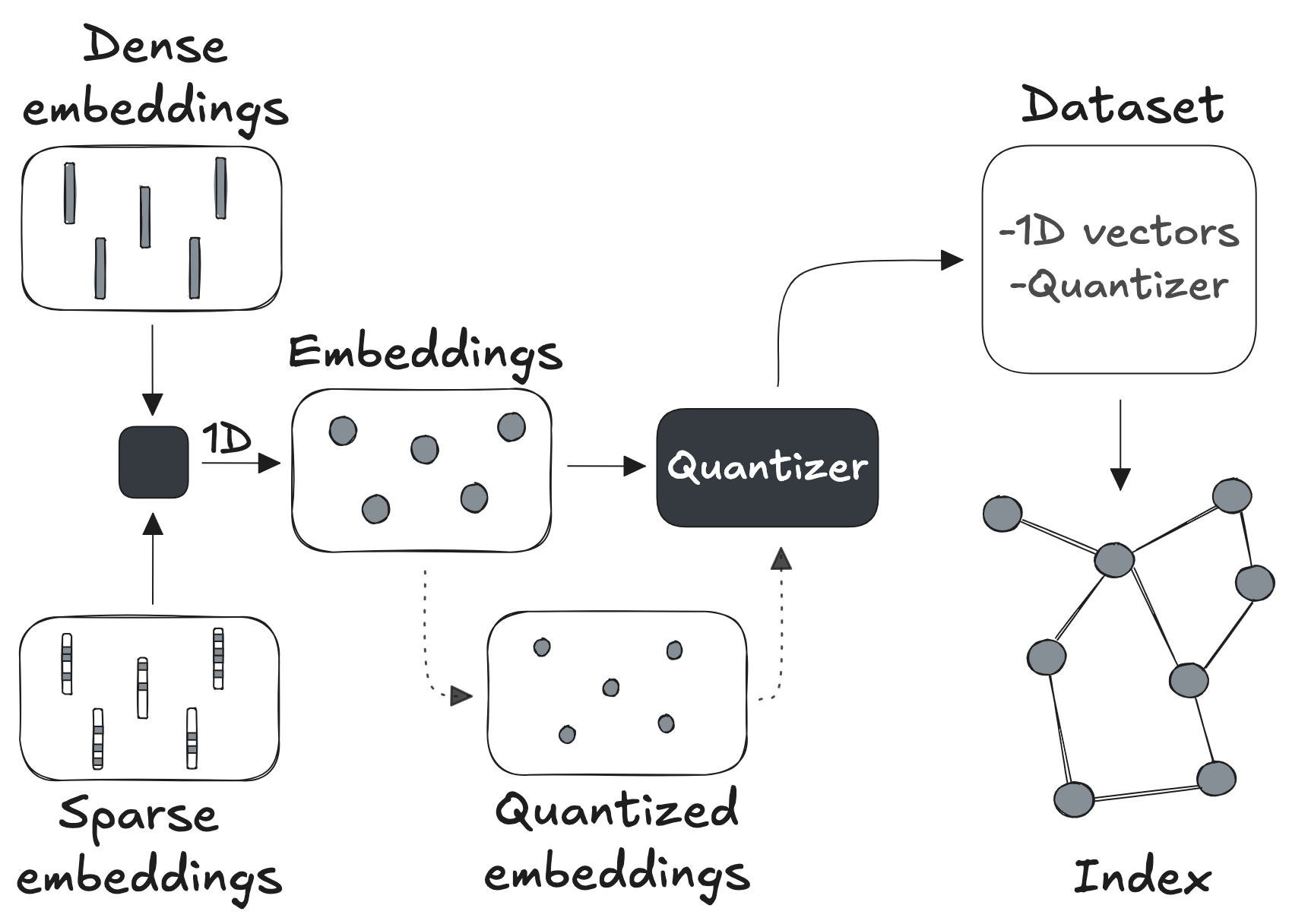}
\captionsetup{hypcap=false}\captionof{figure}{\small Indexing with \kannolo.}\label{fig:indexing}
\end{minipage}\hspace{0.3cm}
\begin{minipage}[t]{0.46\textwidth}
\includegraphics[width=\textwidth]{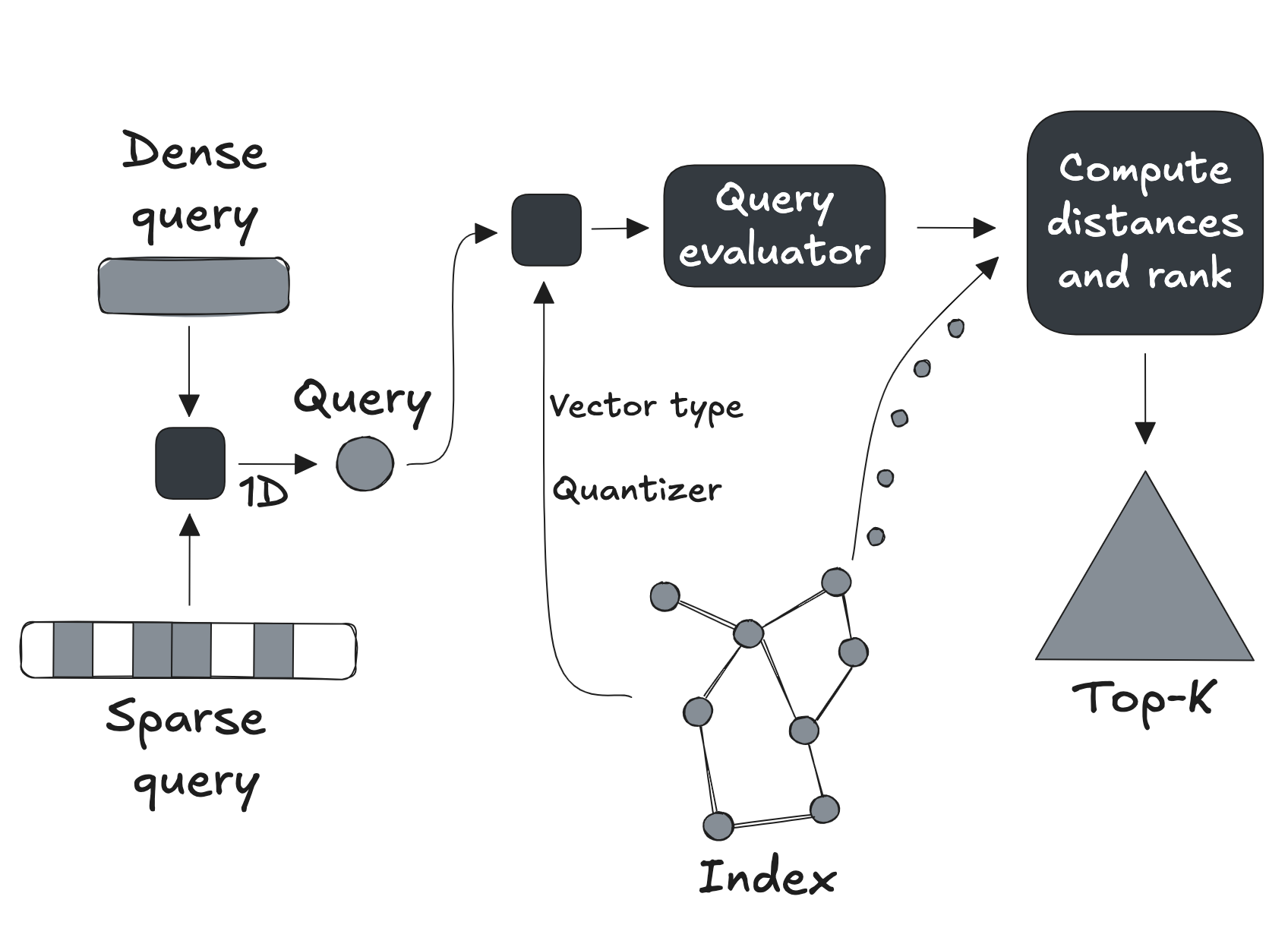}
\captionsetup{hypcap=false}\captionof{figure}{\small Retrieval with \kannolo. \qquad \quad }\label{fig:retrieval}
\end{minipage}
\medbreak
\vspace{1mm}
In \kannolo, the dataset provides access to the data while the index is meant to pinpoint which vector shall be compared with the query. 
The indexing method currently implemented in \kannolo is the Hierarchical Navigable Small World graph (HNSW) \cite{hnsw}, a graph index with a hierarchical links structure that has proved to be a state-of-the-art method both for dense and sparse retrieval. The quantization method that \kannolo implements is the Product Quantization \cite{pq}, which quantizes the data into subspaces independently.

\section{Experiments}

In this section, we evaluate the performance of \kannolo by comparing it to the best ANN libraries implementing the same methods.

\vspace{1mm}
\noindent \textbf{Datasets}.
For the dense domain, we evaluate \kannolo with Euclidean distance on the widely-known \sift dataset \cite{sift}, consisting of vectors of handcrafted visual descriptors of images. Additionally, we test \kannolo with inner product on two sets of embeddings from the \msmarco passages dataset~\cite{msmarco}, produced by the \stard~\cite{star} and \dragon \cite{dragon} models. For sparse retrieval, we use inner product on a sparse version of the \msmarco dataset, where the \splade model generates embeddings \cite{splade}.
For additional details about the tested datasets refer to Table \ref{table1}.

\vspace{-2mm}
\begin{table}
\centering
\caption{Summary of the datasets employed in our experimental evaluation.}\label{table1}
\begin{tabular}{lllrr}
\toprule
\textbf{Dataset Name \ \ } & \textbf{Data Type \ \ } & \multicolumn{1}{c}{\textbf{Measure \ \ }} & \multicolumn{1}{l}{\textbf{Cardinality \ \ }} & \multicolumn{1}{l}{\textbf{Dimensionality}} \\
\midrule
\sift & Dense & L2 & $1$,$000$,$000$ \ \ & $128$ \\ 
\dragon/\stard & Dense & IP & $8$,$841$,$823$ \ \ & $768$ \\ 
\splade & Sparse & IP & $8$,$841$,$823$ \ \ & - \\
\bottomrule
\end{tabular}
\end{table}

\vspace{-1mm}
\noindent \textbf{Competitors}. Dense competitors were selected based on their performance on the wide-recognized benchmark repository ANN benchmarks~\cite{ann_bench}.
%We test \kannolo against state-of-the-art competitors in the sparse and dense domains. 
Specifically, they are: 1) \faiss, Meta's library for ANN search with support for a wide range of ANN methods, including HNSW and PQ; 2) \hnswlib \cite{hnsw}, the original implementation of the HNSW algorithm, part of the nmslib library \cite{nmslib}; 3) \nn \cite{n2}, a lightweight ANN library implementing HNSW. 

\noindent The sparse competitors are the winner submissions of the ``Sparse Track''  at NeurIPS 2023 Big ANN challenge~\cite{annneurips23}, namely \grassrma \cite{grassrma} and \pyann \cite{pyann}. These methods implement HNSW with only some minor optimizations, as \kannolo. 
%However, since the algorithm implemented in \kannolo is a graph index, we test \kannolo against the winning solutions of the ``Sparse Track'' at the 2023 BigANN Challenge \cite{annneurips23} at NeurIPS, which are both graph-based solutions.
%These include \grassrma \footnote{C++ code is publicly available at \url{https://github.com/Leslie-Chung/GrassRMA}.}, a method for dense vectors adapted to sparse vectors, that appears in the BigANN challenge as ``\textsc{sHnsw}'', and \pyann \footnote{C++ code is publicly available at \url{https://github.com/veaaaab/pyanns}.}, the winner solution. 
We leave the comparison with inverted indexes for future work~\cite{seismic}.

\vspace{2mm}
\noindent \textbf{Reproducibility and Hardware Details}.
We implemented \kannolo in Rust (version $1{.}83$) with the compiler’s highest optimization level enabled. We conduct experiments on a server with one Intel i9-9900K CPU clocked at 3.60GHz, with $64$ GiB of RAM. 
%The CPU has $8$ physical cores and $8$ hyper-threaded ones. 
We query the indexes using a single thread.

\subsection{Experimental Results}
In this section, we experimentally show that \kannolo meets both its goals: 

\vspace{1mm}
\noindent \textbf{Modularity}. The tests on both sparse and dense datasets, with plain and product quantizers, were conducted using the same indexing algorithm, which operates independently of the vector representation, quantizer, and similarity measure used. On the other hand, all our competitors have specialized implementations tailored for the vector type or quantization method used; 

\begin{figure}
    \centering
  \includegraphics[width=\textwidth]{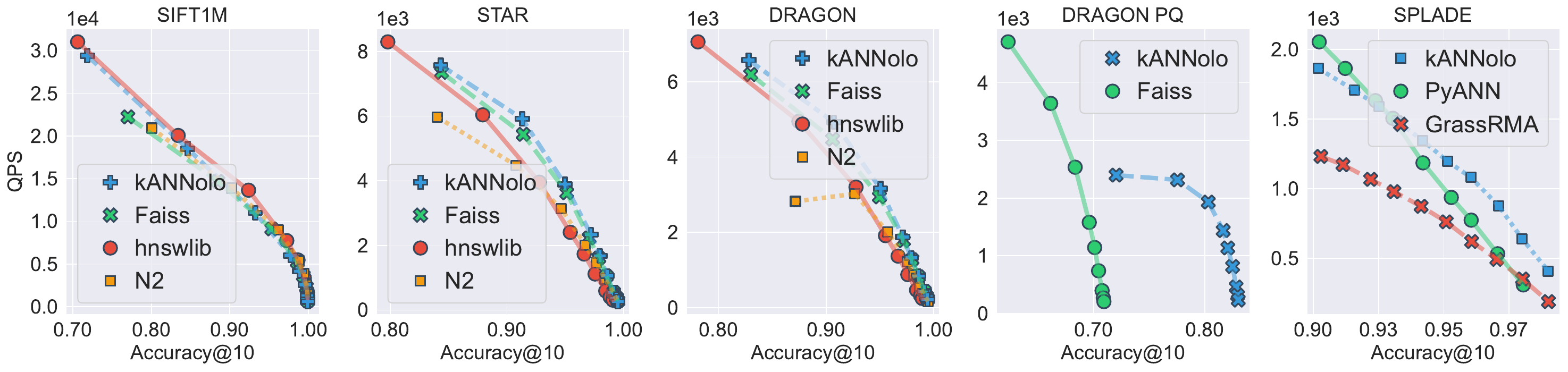}
    \caption{\small Accuracy vs. queries per second. Left to right: Dense vectors, no quantization (1-3), dense vectors with product quantization (4), sparse vectors, no quantization (5).}
    \label{fig:benchmarks}
\end{figure}

\vspace{1mm}
\noindent \textbf{Performance}: \kannolo achieves an improvement over state-of-the-art libraries in terms of accuracy-speed trade-off (Figure \ref{fig:benchmarks}). On dense data, \kannolo is competitive with the state-of-the-art libraries on \sift dataset, while it outperforms its competitors on \msmarco. In particular, \kannolo is up to $11.1\times$ faster than \faiss as it builds the HNSW graph on the original representations of the vectors rather than on the quantized ones.
%both with plain and product quantizers (up to 11.1$\times$ faster) on a classical Information Retrieval dataset like \msmarco. 
%where the high dimensionality and the vectors' distribution make the task of finding the nearest neighbors harder.
On sparse vectors \kannolo outperforms both competitors, with up to 2.1$\times$ speedup.

\vspace{-1mm}
\section{Conclusions and Future Work}

We introduced \kannolo, a Rust-based library for approximate nearest neighbors search. It is designed to ease the prototyping and development of new ANN algorithms. Our extensive benchmarking on several public dense/sparse datasets shows that \kannolo achieves genuine modularity while also delivering state-of-the-art performance against well-known competitors. These results confirm that \kannolo is a valuable tool for effectively advancing ANN research.

As future work, we plan to extend \kannolo with additional indexing and quantization methods, thereby providing researchers with a broader experimental playground. 
%More generally, we will keep \kannolo up to date to ensure its continued relevance and utility in the evolving field of ANN search.

\section*{Acknowledgments}

This work was partially supported by the Horizon Europe RIA ``Extreme Food Risk Analytics'' (EFRA), grant agreement n. 101093026, by the PNRR - M4C2 - Investimento 1.3, Partenariato Esteso PE00000013 - ``FAIR - Future Artificial Intelligence Research'' - Spoke 1 ``Human-centered AI'' funded by the European Commission under the NextGeneration EU program, by the PNRR ECS00000017 Tuscany Health Ecosystem Spoke 6 ``Precision medicine \& personalized healthcare'' funded by the European Commission under the NextGeneration EU programme, by the MUR-PRIN 2017 ``Algorithms, Data Structures and Combinatorics for Machine Learning'', grant agreement n. 2017K7XPAN\_003, and by the MUR-PRIN 2022 ``Algorithmic Problems and Machine Learning'', grant agreement n. 20229BCXNW.

\bibliographystyle{splncs04}
\bibliography{references.bib}

\end{document}

%% file: macros.tex
\newcommand{\msmarco}{\textsc{Ms Marco}\xspace}
\newcommand{\splade}{\textsc{Splade}\xspace}
\newcommand{\sift}{\textsc{Sift1M}\xspace}
\newcommand{\dragon}{\textsc{Dragon}\xspace}
\newcommand{\stard}{\textsc{Star}\xspace}
\newcommand{\kannolo}{\texttt{kANN}olo\xspace}
\newcommand{\grassrma}{\textsc{GrassRMA}\xspace}
\newcommand{\pyann}{\textsc{PyAnn}\xspace}
\newcommand{\faiss}{\textsc{Faiss}\xspace}
\newcommand{\hnswlib}{\textsc{hnswlib}\xspace}
\newcommand{\nn}{\textsc{N2}\xspace}